\documentstyle[a4p,12pt,epsfig,subfigure,array,cite,pennames]{article}
\newcommand{\defaultstretch}
    {\small\normalsize}
\newcommand{\inmath}[1] {\ifmmode#1\else$#1$\fi}
\newcommand{\definmath}[2] {\def#1{\ifmmode#2\else$#2$\fi}}
\def\qq{\ifmmode {{\mathrm q\bar{\mathrm q}}}
    \else {${\mathrm q\bar{\mathrm q}}$} \fi}
\definmath{\Pq}      {\mathrm{q}}
\definmath{\Paq}  {\overline{\mathrm{q}}}
\def\K{\ifmmode {{\mathrm K}} \else {${\mathrm K}$}\fi}
\def\Zz{\ifmmode {{\mathrm Z}^0} \else {${\mathrm Z}^0$}\fi}
\def\dEdx{\ifmmode {{\mathrm d}E/{\mathrm d}x}\else
                   {${\mathrm d}E/{\mathrm d}x$}\fi }
\def\dedx{\ifmmode { {\mathrm d}E/{\mathrm d}x}\else
                     {${\mathrm d}E/{\mathrm d}x$}\fi }
\def\pipi{\ifmmode {\pi^+\pi^-}\else {$\pi^+\pi^-$}\fi}
\def\KK{\ifmmode {{\mathrm K}^+{\mathrm K}^-}\else
                  {K$^+$K$^-$}\fi}
\def\fzz{\ifmmode {{\mathrm f}_0(980)}\else
                  {f$_0(980)$}\fi }
\def\fz{\ifmmode {{\mathrm f}_0}\else
                  {f$_0$}\fi }
\def\ftwo{\ifmmode {{\mathrm f}_2(1270)}\else
                  {f$_2(1270)$}\fi }
\def\ft{\ifmmode {{\mathrm f}_2}\else
                  {f$_2$}\fi }
\def\ph{\ifmmode {\phi}\else
                 {$\phi$}\fi }
\def\KKbar{\ifmmode {{\mathrm K}\overline {\mathrm K}}\else
                   {K$\overline {\mathrm K}$}\fi }
\newcommand{\qqbar}  {\Pq\Paq}

\def\costhstar{\ifmmode {\cos\theta^*} \else $\cos\theta^*$ \fi }

\newcommand{\likepipi}{$\pi^\pm\pi^\pm$}
\newcommand{\likeKK}{K$^\pm$K$^\pm$}

\newcommand{\fourq}{qq$\bar{\mathrm q}\bar{\mathrm q}$}
\newcommand{\xp}{$x_p$}

\begin{document}
\begin{titlepage}
\begin{center}{\large   EUROPEAN LABORATORY FOR PARTICLE PHYSICS}
\end{center}\bigskip

\begin{flushright}
       CERN-EP/98-010   \\ 23rd January 1998 
\end{flushright}
\bigskip\bigskip\bigskip\bigskip\bigskip
\begin{center}{\huge\bf\boldmath
Production of \fzz, \ftwo\ and 
$\phi(1020)$ in hadronic \Zz\ decay
}\end{center}\bigskip\bigskip
\begin{center}{\LARGE The OPAL Collaboration
}\end{center}\bigskip\bigskip
\bigskip\begin{center}{\large  Abstract}\end{center}
Inclusive production of the \fzz, \ftwo\ and $\phi(1020)$ 
resonances has been studied in
a sample of 4.3 million hadronic \Zz\ decays from the OPAL experiment
at LEP. A coupled channel analysis has been used for the \fz\
in simultaneous fits to the resonances in inclusive \pipi\ and \KK\ mass 
spectra.
Fragmentation functions are reported for the three states. 
Total inclusive rates are measured to 
be $0.141 \pm 0.007 \pm 0.011$ \fz, $0.155\pm0.011\pm0.018$ \ft\ 
and $0.091\pm0.002\pm0.003$ $\phi$ mesons per hadronic \Zz\ decay.
The production properties of the \fz, including those in three-jet 
events, are compared with those of the \ft\ and $\phi$, and with the 
Lund string model of hadron production. All measurements are 
consistent with the 
hypothesis that the \fzz\ is a conventional \qqbar\ scalar meson.
\bigskip\bigskip\bigskip\bigskip
\bigskip
%\begin{center}
%{\large {\bf Comments} to George.Lafferty@man.ac.uk 
%by Thursday 22nd January 98}
%\end{center}
\bigskip
\bigskip\bigskip\bigskip
\begin{center}{\large
Submitted to European Physical Journal C: Particles and Fields
}\end{center}
\end{titlepage}

%\begin{center}
%{\bf Editorial Board}: Richard Hawkings, Richard Hemingway, Harold Ogren,
%Andr\'e Turcot
%\end{center}
%\end{titlepage}

\begin{center}{\Large        The OPAL Collaboration
}\end{center}\bigskip
\begin{center}{
%begin authorlist
K.\thinspace Ackerstaff$^{  8}$,
G.\thinspace Alexander$^{ 23}$,
J.\thinspace Allison$^{ 16}$,
N.\thinspace Altekamp$^{  5}$,
K.J.\thinspace Anderson$^{  9}$,
S.\thinspace Anderson$^{ 12}$,
S.\thinspace Arcelli$^{  2}$,
S.\thinspace Asai$^{ 24}$,
S.F.\thinspace Ashby$^{  1}$,
D.\thinspace Axen$^{ 29}$
% ...
%end authorlist
}\end{center}\bigskip
\bigskip
%begin institutes
$^{  1}$School of Physics and Astronomy, University of Birmingham,
Birmingham B15 2TT, UK
\newline
$^{  2}$Dipartimento di Fisica dell' Universit\`a di Bologna and INFN,
I-40126 Bologna, Italy
\newline
% ...
$^{  5}$Cavendish Laboratory, Cambridge CB3 0HE, UK
\newline
% ...
$^{  8}$CERN, European Organisation for Particle Physics,
CH-1211 Geneva 23, Switzerland
\newline
$^{  9}$Enrico Fermi Institute and Department of Physics,
University of Chicago, Chicago IL 60637, USA
\newline
% ...
$^{ 12}$Indiana University, Department of Physics,
Swain Hall West 117, Bloomington IN 47405, USA
\newline
% ...
$^{ 16}$Department of Physics, Schuster Laboratory, The University,
Manchester M13 9PL, UK
\newline
% ...
$^{ 23}$Department of Physics and Astronomy, Tel Aviv University,
Tel Aviv 69978, Israel
\newline
$^{ 24}$International Centre for Elementary Particle Physics and
Department of Physics, University of Tokyo, Tokyo 113, and
Kobe University, Kobe 657, Japan
\newline
% ...
$^{ 29}$University of British Columbia, Department of Physics,
Vancouver BC V6T 1Z1, Canada
\newline
% ...
\newline
%end notes

\newpage
\section{Introduction}
\label{sec-intro}

Inclusive production of mesons and baryons in Z$^0$ decay has 
been studied 
extensively~\cite{Bohrer}, and the results have provided valuable
input to the theory and phenomenology of parton 
hadronization.
Production of the \fzz\ is of particular interest because
although it is well established experimentally as a scalar
(${\mathrm J}^{\mathrm {PC}}=0^{++}$) state, its precise nature has long 
been uncertain. Indeed,
there is still vigorous debate~\cite{Pennington, PDG} about the identity 
of the mesons which comprise the lowest lying scalar nonet of flavour
SU(3).
Compared to the expectations for a conventional meson at a mass of 
around 1~GeV, 
the \fz\ has a markedly 
small total width, a relatively large coupling to 
\KKbar, and a partial width to $\gamma\gamma$ which is an order of magnitude
below theoretical expectation. 

A number of suggestions have been made as to the nature of the \fz. 
Jaffe and Johnson~\cite{Jaffe} performed a bag model calculation 
to suggest that 
it could be a ``cryptoexotic'' \fourq\ state. Weinstein and 
Isgur~\cite{Weinstein}, using a potential model of \fourq\ states, 
showed that the \fz\ could be explained as a loosely bound \KKbar\ system, 
a so-called \KKbar\ molecule. Gribov~\cite{Gribov, Close} has 
proposed a theory of 
confinement in QCD, in which the \fz\ plays the role of a novel 
``vacuum scalar'' state, a bound state of a 
quark and antiquark with negative kinetic energy, interacting 
repulsively to give a state of positive total energy. 
Ishida et al.~\cite{Ishida} have proposed an 
interpretation as a
hybrid meson with a massive constituent gluon, while
a scalar glueball has been suggested by Robson~\cite{Robson}. 
An analysis by Close and Amsler~\cite{Closeglue} of the 
scalar states suggests that the most likely candidate for the
lowest-lying scalar glueball is the f$_0(1500)$; in this model, 
the \fzz\ is a left-over state which cannot be accommodated in 
the scalar nonet. Lattice QCD calculations~\cite{Lattice} also find 
the lightest 
scalar glueball mass to be around 1.5~GeV so that the glueball
interpretation of the \fz\ is now out of favour.
T\"ornqvist~\cite{Tornqvist}, using a 
unitarized quark model to 
analyse all of the identified states in the $0^{++}$ sector, has concluded
however 
that the \fz\ can fit in as a conventional \qqbar\ meson in the
lowest $0^{++}$ multiplet. Similarly, an analysis by Zou and Bugg~\cite{Bugg}
of all available high-statistics \pipi\ 
and \KKbar\ scattering data concluded 
that the \fz\ could be interpreted as a meson, and recent work by
Anisovich and Sarantsev~\cite{Anisovich} has come to the same
conclusion.

One aim of the present study is to look for features of \fz\ 
production in Z$^0$ decay which may help to elucidate 
its nature. The vacuum scalar states of Gribov's theory are expected 
to be compact objects with 
distinctive production properties, and suggestions have been 
made by~Close et al.~\cite{Close} for a number of experimental tests. 
In Z$^0$ decay, the signature would be a relatively larger yield in low 
multiplicity events and in events where the \fz\ is isolated in rapidity. 
If, on the other hand, the \fz\ has a significant gluonic content, its 
production could be enhanced in gluon jets. 
In contrast, if it is principally a 
conventional meson, its production properties may be unremarkable when 
compared to those of other similar states. There are presently no
predictions for production rates of \fourq\ states or \KKbar\ molecules
in Z$^0$ decay.

Two approaches are taken in the present analysis: a comparison of the 
features of \fz\ production with those of two neutral isoscalar mesons, 
and a comparison with the JETSET~7.4~\cite{JETSET} implementation of the 
Lund string model of hadronization~\cite{string},
within which the \fz\ is treated as a conventional meson. 
The features of \fz\ production are measured along with those of the 
f$_2(1270)$ and the $\phi(1020)$ mesons. The latter is close in mass 
to the \fz,
while the former is an established meson with the \qqbar\ in a
$^3{\mathrm P}_2$ state. In the conventional meson interpretation, 
the \fz\ would be a $^3{\mathrm P}_0$ state. Since the mass 
difference is relatively small, the \fz\ and f$_2$ may then be expected to 
have similar production properties.
The Lund string model, as implemented in the 
JETSET~7.4 Monte Carlo program, is highly successful in
describing many features of hadronic Z$^0$ decay events. These features 
include global event properties related to the perturbative phase of the 
initial parton shower, as well as details of the production of many 
different species of hadrons during the nonperturbative hadronization 
phase. In short, the model provides a well-tested and reliable picture 
of particle production
in hadronic Z$^0$ decays~\cite{toprev}. Within the JETSET model, 
the \fz\ is treated as a scalar meson composed of u$\overline {\mathrm u}$ 
and d$\overline {\mathrm d}$ pairs, and the relative production rates of
$^3{\mathrm P}_0$ and $^3{\mathrm P}_2$ mesons are determined by a 
variable parameter.
A comparison of the data and the results of JETSET could therefore
provide information on the nature of the \fz. 
 
The only reported measurements of the \fz\ and \ft\ at LEP were made by 
DELPHI as part of a general study of light resonances~\cite{Dlight}
using relatively 
low statistics. The fragmentation functions and total rates were reported
over restricted ranges of momentum. 
In a similar study of resonance production
by ALEPH~\cite{Arho}, the \fz\ was included in fits to \pipi\ 
mass spectra, but no rates were reported. Because the \fz\ 
has a significant coupling to \KKbar\, and peaks 
below threshold for this channel, its \pipi\ mass spectrum will not 
in general
exhibit the simple Breit-Wigner shape assumed in these two previous analyses. 
In the present study, a coupled-channel analysis is done, 
including $\pi^+\pi^-$ and \KK\ data, in which proper account is taken
of the opening of the \KKbar\ channel in \fz\ decay. 

%The coupling of the \fz\ to \KKbar\ leads to a
%low mass enhancement in the \KzsKzs\ channel, the intensity of which is an
%important input to studies of Bose-Einstein correlations~\cite{BECs}.
%A reliable measurement of \fz\ production in the LEP data would enable
%proper account to be taken of this effect.
%One outcome of the present analysis is
%a prediction of the cross section for the component of the low mass 
%\KzsKzs\ enhancement arising from \fzz\ production. 

\section{The OPAL detector and data samples}
\label{sec-OPAL}

The OPAL detector is described in~\cite{OPALdet}. For the 
present analysis, the most important components were the central
tracking chambers which consist of two layers of
silicon microvertex detectors~\cite{silicon}, a high-precision 
vertex drift chamber,
a large-volume jet chamber, and a set of drift chambers (the $z$-chambers)
which measure the coordinates of tracks along the direction of
the beam. The OPAL coordinate
system is defined with the $z$-axis following the electron beam direction; 
the polar angle~$\theta$~is
defined relative to this axis, and $r$ and $\phi$ are the usual
cylindrical polar coordinates.
The central chambers lie within a homogeneous axial magnetic field
of $0.435$~T. Charged particle tracking is
possible over the range $|\cos \theta|<0.98$ for the full range of 
azimuthal angles.
The OPAL jet chamber is capable of measuring specific energy loss, \dedx,
with a resolution, $\sigma(\dedx)/(\dedx)$, of 3.5\% for 
well-reconstructed, high-momentum 
tracks in hadronic events~\cite{DEDX}.

The present analysis used the full OPAL sample of 4.3 million hadronic 
Z$^0$ decays recorded at LEP 1 between 1990 and 1995. To correct for 
losses due to the acceptance and efficiency of the experiment 
and the selection procedures, and also to provide signal and 
background shapes for fits to the data mass spectra, 6 million 
Monte Carlo events were used, which had been generated using
JETSET 7.4 and processed
through a full simulation of the experiment~\cite{GOPAL} and the 
data reconstruction and analysis. This `detector-level' Monte Carlo
sample was also used for comparison with various features of 
the experimental data.
The JETSET version was tuned~\cite{JTtune} 
using OPAL data on event shape distributions, fragmentation
functions of $\pi^\pm$, K$^\pm$, p/${\overline {\mathrm p}}$
and $\Lambda$, and LEP data on total inclusive
multiplicities for 26 identified hadron species.

A detailed description of the selection of hadronic Z$^0$ decay events 
in OPAL is given in~\cite{TKMH}. For the present analysis, tracks 
in the selected events were
required to have: a minimum momentum transverse to the beam direction
of 150~MeV/$c$; a maximum momentum of $1.07 \times E_{\mathrm {beam}}$,
based on the momentum resolution of the detector; a distance
of closest approach to the interaction point less than 5~cm in the 
plane orthogonal
to the beam direction, and the corresponding
distance along the beam direction less than 40~cm; 
a first measured point within a radius of 75~cm 
from the vertex; and at least 20 hits available for measurement of
specific energy loss, \dedx.
 
Kaons and pions were identified using the \dedx\ measurements.
For each track, a $\chi^2$ probability (weight) was formed for each of the
stable particle hypotheses (e, $\mu$, $\pi$, K and p). A track was 
identified as a pion or a kaon if
the appropriate weight was above 5\% and was larger than the weight for
each of the other stable hadron hypotheses. 
Between momenta of 0.8 and 2.0~GeV/$c$, 
the $\pi$, K and p bands overlap in \dedx, leading to considerable
ambiguity among hypotheses. Therefore, no tracks were identified as kaons 
in this momentum range although, since most particles are pions, 
pion identification was still allowed. 

With the event and track selection cuts previously described,
\fz\ and \ft\ decaying via \pipi\ were identified with an 
efficiency of around 40\% over the whole momentum range. 
The $\phi \rightarrow \KK$ efficiency typically varied between $15$ 
and $20\%$, although
in the $\phi$ momentum range from 1.4 to 3.6~GeV/$c$ it
fell below $5\%$ because of the \dedx\ cross-over region. 
The mass resolution at 1~GeV
varied with momentum, from 15 to 20~MeV for \pipi\ and
from 2.5 to 4.5~MeV for \KK.

\section{Data analysis}
\label{analysis}

\subsection{Inclusive two-particle mass spectra}
\label{sec-massfits}

For simultaneous fits to \pipi\ and \KK\ mass spectra in
both the real and simulated data samples,
spectra were formed for inclusive \pipi, \likepipi\ and \KK\ 
% and \likeKK\ 
systems in bins of each of the following variables:
\begin{itemize}
\item{scaled momentum, \xp ($=p/E_{\mathrm {beam}}$), of the 
two-particle systems
(nine bins)}
\item{rapidity, $y$, of the two-particle systems with respect to the event 
thrust axis (six bins)}
\item{rapidity gap, $\Delta y$, between the two-particle systems and the 
closest single charged particle (six bins)}
\item{multiplicity, $n_{\mathrm {ch}}$, of charged tracks in the event 
(six bins)}
\end{itemize}
The last three of these variables were chosen specifically for tests 
of the nature of the \fz, as discussed in section 1.
To account for a large part of the combinatorial backgrounds in \pipi, 
mass spectra were formed by subtracting the \likepipi\ 
spectra from those for \pipi. Because the combinatorial backgrounds in
\KK\ were relatively small, and Bose-Einstein correlations could affect
the \likeKK\ spectra near threshold, no subtraction was 
done for the \KK\ mass spectra. 
For the Monte Carlo sample, separate
spectra were also made for the most important states contributing
to the mass spectra, using information on the origin of each track
at the generator level.

\subsection{Selection of three-jet events}
\label{jetfinding}

To investigate possible differences between production in quark 
and gluon jets, the Durham jet finder~\cite{Durham} was used to
identify a sample of three-jet events. The
cut-off value $y_{\mathrm {cut}}$ was set to $0.005$, and the jet-finding 
was done using charged tracks. For each candidate three-jet event,
the angle between the two 
lowest-energy jets was required to be greater than $30^\circ$, 
and in order to ensure well-reconstructed, planar events, the 
sum of the interjet angles was required to be larger than 
$358^\circ$. The jet energies were then reconstructed using the interjet 
angles, assuming massless kinematics.
%~\cite{massless}. 
Each jet was required to contain at least two charged particles and
more than 5 GeV of energy. With these cuts, $24\%$ of all events
were selected as three-jet events.
Monte Carlo studies
%~\cite{Bonn} 
have shown that in such events, $95\%$ of
the highest-energy jets are due to quarks, while
the lowest-energy jet is a gluon jet with an approximately $80\%$ probability.
Mass spectra were formed, using only tracks
assigned to the same jet, in three bins of $E/E_{\mathrm {jet}}$
where $E$ is the energy of the two-particle system and 
$E_{\mathrm {jet}}$ is the energy of the jet. 

\subsection{Fit procedures}

For each bin of the above kinematic variables, the
\pipi\ and \KK\ mass spectra 
were fitted simultaneously, using a minimum $\chi^2$ fit,
to a sum of contributions 
given by:

\begin{equation}
f(m_\pipi) = a_{\fz}^{\pipi} |A_{\fz}(m_{\pi\pi)}|^2+
a_{\ft}{\mathrm {BW}}_{\ft}
+a_{\mathrm {bgd}}^\pipi B_{\pipi}
\label{fitpipi}
\end{equation}

\begin{equation}
f(m_\KK) = a_{\fz}^{\KK} |A_{\fz}(m_{\mathrm {KK}})|^2
+a_{\ph}{\mathrm {BW}}_{\ph}
+a_{\mathrm {bgd}}^{\KK} B_{\KK}
\label{fitKK}
\end{equation}

\noindent
In equations~(\ref{fitpipi}) and (\ref{fitKK}), the $a$ terms represent
the intensities to be fitted, $A_\fz$ is the amplitude for \fzz,
BW are Breit-Wigner functions, and the $B$ represent 
background functions. The $a_{\fz}$ are related by
$$a_{\fz}^{\KK} = 0.75{e_{\KK}\over e_{\pipi}} a_{\fz}^{\pipi}$$
where ${e_{\KK}/e_{\pipi}}$ is the ratio 
of the efficiency to reconstruct \fz\ in
\KK\ relative to \pipi\ (which varies with the bin of the 
kinematic variable), and the factor 0.75 comes from  
Clebsch-Gordan coefficients.
Following Flatt\'e~\cite{flatte}, the coupled-channel 
amplitudes for \fzz\ decay via $\pi\pi$ and KK were taken to be:

\begin{equation}
A_{\fz}(m_{\pi\pi}) = {m_0\sqrt{\Gamma_{\pi\pi}} \over
m_0^2-m_{\pi\pi}^2-im_0\left(\Gamma_{\pi\pi}+\Gamma_{\mathrm {KK}}\right)}
\label{flattepipi}
\end{equation}

\begin{equation}
A_{\mathrm {\fz}}(m_{\mathrm {KK}}) = 
{m_0\sqrt{\Gamma_{\mathrm {KK}}} \over
m_0^2-m_{\mathrm {KK}}^2-im_0\left(\Gamma_{\pi\pi}
+\Gamma_{\mathrm {KK}}\right)}
\label{flatteKK}
\end{equation}

\noindent
Here $m_0$ is the resonance mass, and the partial widths $\Gamma$
are related to the coupling constants $g$ via
$$
\Gamma_{\pi\pi} = g_\pi\sqrt{{m_{\pi\pi}^2\over 4}-m^2_\pi} 
\hspace{1.cm}{\mathrm {and}}\hspace{1.cm}
\Gamma_{\mathrm {KK}} = g_{\mathrm K}\sqrt{{m_{\mathrm {KK}}^2
\over 4}-m^2_{\mathrm K}}
$$
The fact that $\Gamma_{\mathrm {KK}}$ is imaginary below KK threshold leads
to distortion of the $\pi\pi$ mass spectrum from a simple Breit-Wigner
shape. Interference between \fz\ and \pipi\ backgrounds in Z$^0$ decay 
should be negligible and was not included in the fits. In
determining the meson rates, the possible presence of other 
f-meson resonances coupling to \pipi\ has been neglected. 

The experimental mass resolution was treated differently 
for the three states. 
The \fz\ intensities derived from equation (3) were folded with the 
detector resolution
separately for each bin of \xp, $y$, $\Delta y$,
$n_{\mathrm {ch}}$ and $E/E_{\mathrm {jet}}$.
For the broad \ft, a relativistic D-wave Breit-Wigner was used, with mass
and width fixed to the Particle Data Group (PDG) values~\cite{PDG}. 
For the $\phi$, where experimental mass resolution is particularly
important, the shapes for the fits were taken from the detector-level 
output of the simulation,
accounting automatically for variations of the
mass resolution. The background shapes for the fits were also
taken from the detector-level Monte Carlo: for \pipi, all pairs
identified as \pipi\ were taken, except those from the \fz\ and
\ft; and for the \KK, all identified combinations were used
except those due to \fz\ and $\phi$. In the fits,
the background shapes were given additional freedom, with
the \pipi\ contribution multiplied by a term 
$(m_{\pipi}^{\alpha_1} + \alpha_2 m_{\pipi}^{\alpha_3})$
and the \KK\ shape by $(1+\alpha_4 m_{\KK})$, with the $\alpha$
being variable parameters. 
The $\rho^0$ resonance in the JETSET simulation is 
truncated at $\pm 2.5\Gamma$; to allow for this, the $\rho^0$
intensity above 1.15~GeV was taken as a straight-line 
extrapolation from the contribution at 1.15~GeV falling to 
zero at 1.5~GeV.

\begin{figure}[htbp]
\begin{center}
%\mbox{\epsfig{file=mass.eps,height=17cm,width=17cm}}
\mbox{\epsfig{file=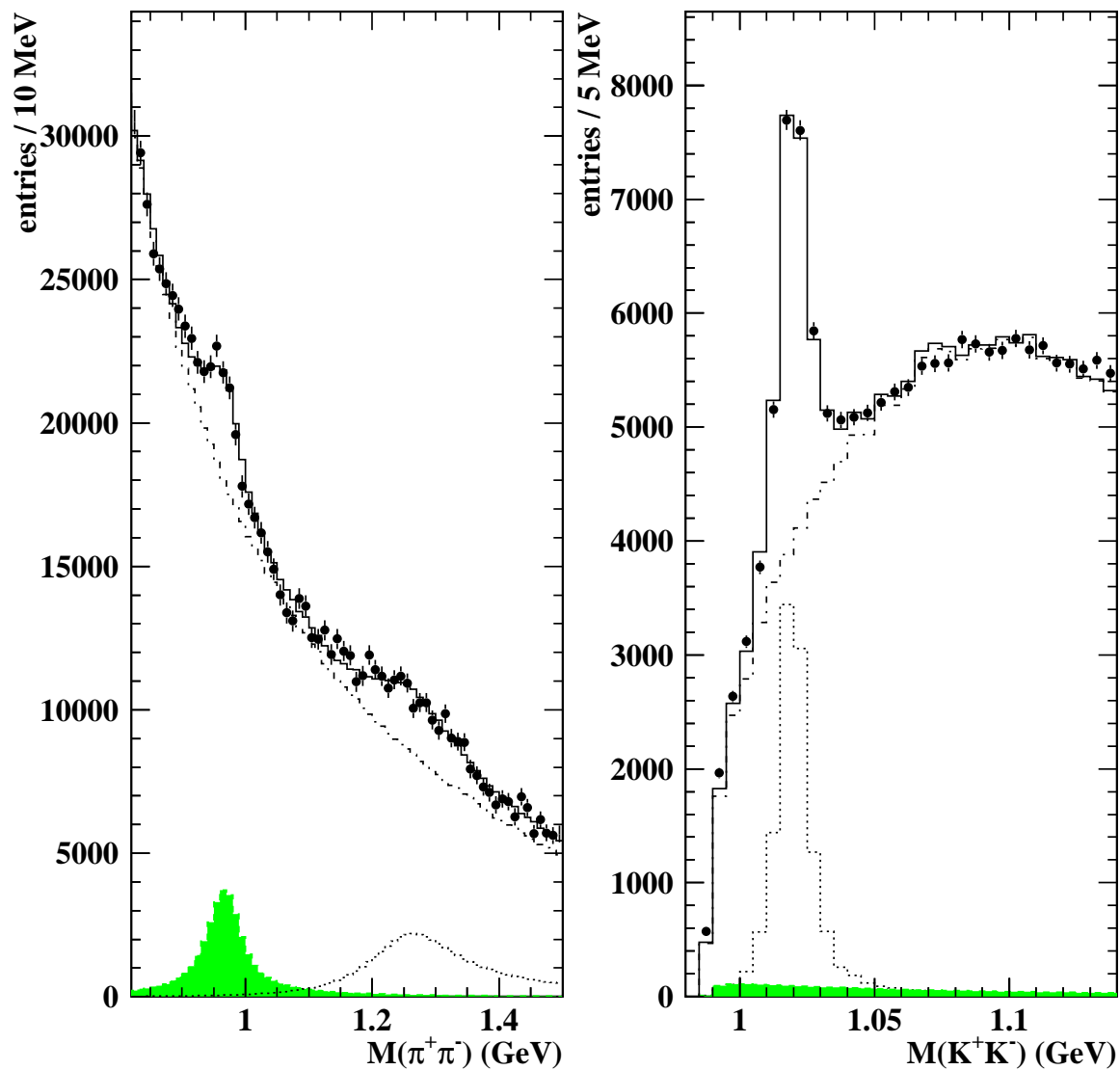,height=17cm,width=17cm}}
\end{center}
\caption{Sum of the fitted mass spectra of \pipi\ 
(with like-sign spectra subtracted)
and \KK\ for $x_p > 0.14$
with the results of the fits described in the text. The points show 
the data, and the solid histograms give the results of the fits. 
The shaded histograms
show the \fz\ contributions, the dotted histograms give the \ft\
and $\phi$
contributions, and the dot-dash histograms show the fitted backgrounds.}
\label{mass}
\end{figure}

The mass ranges of the fits were 0.82 to 1.5~GeV for \pipi,
and from threshold to 1.18~GeV for \KK, these limits being chosen
mainly to avoid the $\rho^0$ peak region in \pipi\ and the region
of its reflection in \KK\ (caused by particle misidentification).
Figure~\ref{mass} shows as an example the 
sum of the fitted mass spectra over the range $x_p > 0.14$, 
indicating clearly the signals due to
the three resonances. 

\begin{figure}[htbp]
\begin{center}
%\mbox{\epsfig{file=xpbins.eps,height=17cm,width=17cm}}
\mbox{\epsfig{file=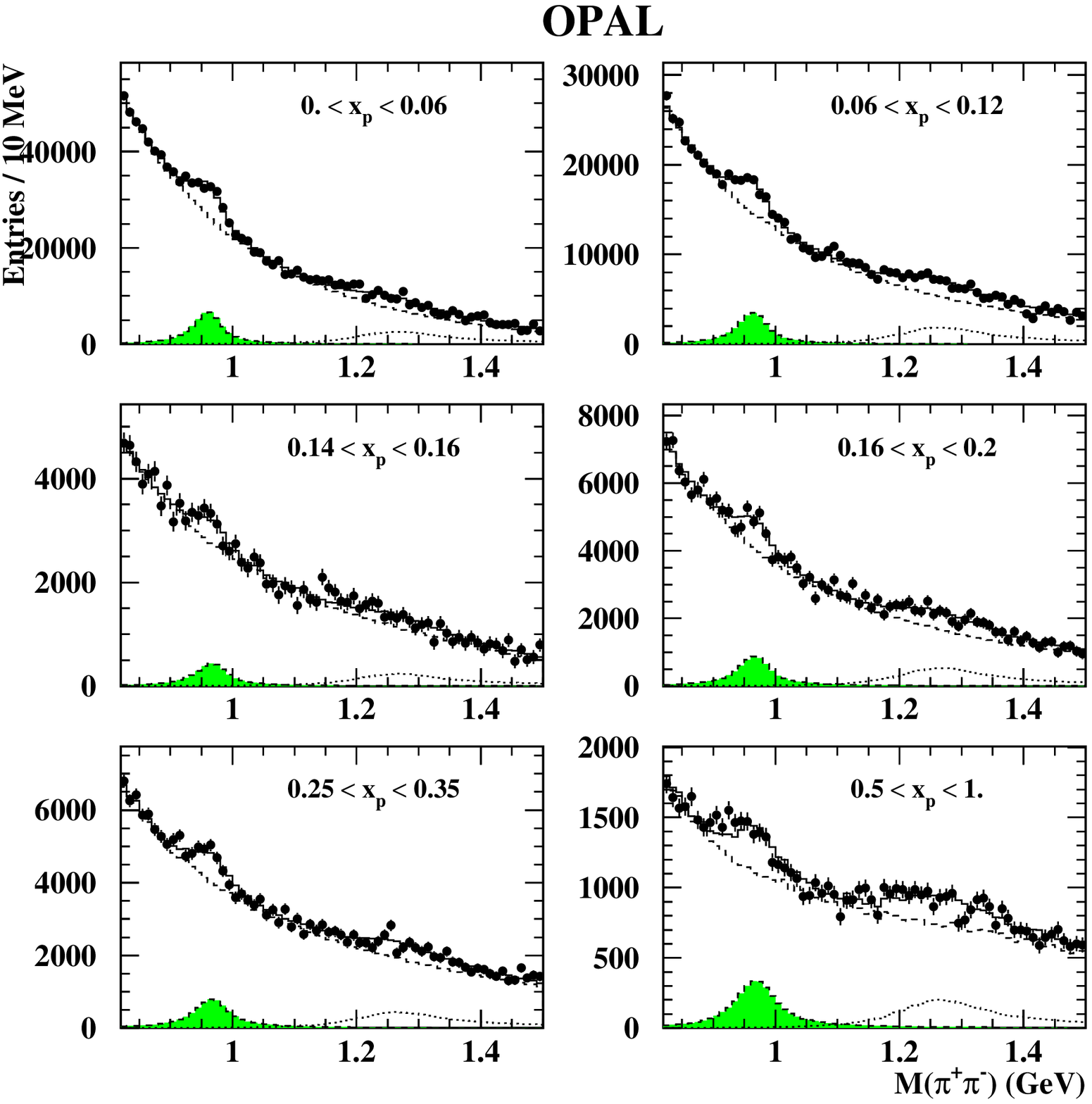,height=17cm,width=17cm}}
\end{center}
\caption{Mass spectra of \pipi\ (with like-sign spectra subtracted)
for 6 bins of $x_p$
with the results of the fits described in the text. The points show 
the data, and the solid histograms give the results of the fits. 
The shaded histograms
show the \fz\ contributions, the dotted histograms show the \ft\
contributions, and the dot-dash histograms show the fitted backgrounds.}
\label{xpbins}
\end{figure}

\begin{figure}[htbp]
\begin{center}
%\mbox{\epsfig{file=xpbinsphi.eps,height=17cm,width=17cm}}
\mbox{\epsfig{file=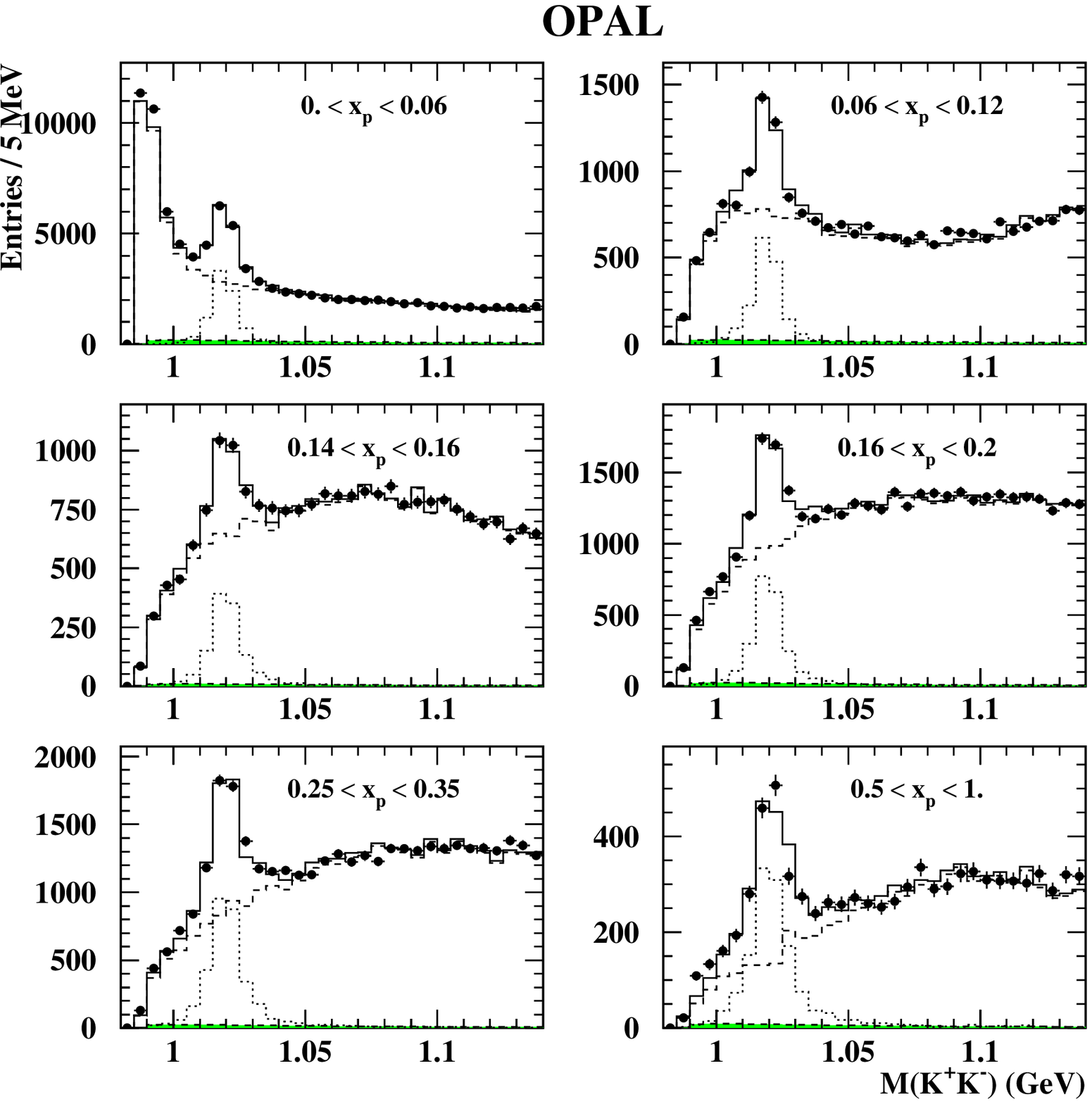,height=17cm,width=17cm}}
\end{center}
\caption{Mass spectra of \KK\  
for 6 bins of $x_p$
with the results of the fits described in the text. The points show 
the data, and the solid histograms give the results of the fits. 
The shaded histograms
show the \fz\ contributions, the dotted histograms show the $\phi$
contributions, and the dot-dash histograms show the fitted backgrounds.}
\label{xpbinsphi}
\end{figure}

In fits to the total mass spectra for $x_p > 0.14$
with the \fz\ parameters allowed to vary, the
mass was well-constrained while the couplings $g$ were poorly determined:
$m_0 = 0.957\pm0.006$~GeV, $g_\pi = 0.09\pm0.40$ and 
$g_{\mathrm K}= 0.97\pm0.82$. This is essentially due to 
the limited mass resolution and the large backgrounds in
the \pipi\ channel; the \KK\ data are relatively insensitive
to the \fz\ parameters.
The fitted mass value is in good agreement
with that obtained in the high-statistics analysis made by 
Zou and Bugg~\cite{Bugg} of $\pi\pi-\KKbar$ elastic scattering phase 
shifts and $\pi\pi$ and \KKbar\ mass spectra from     
central production in pp collisions. The $m_0$ and
$g$ values were therefore fixed to the Zou and Bugg values,
$m_0 = 0.9535$~GeV, $g_\pi$=0.111 and $g_{\mathrm K}=0.423$. These 
couplings correspond
to a branching ratio, BR($\fz \rightarrow \pi\pi) = 0.80$, in agreement
with the PDG value. Nevertheless,
there is much experimental and theoretical uncertainty in the
resonance parameters appropriate for the \fz. Indeed the PDG quotes
the total width as 40 to 100~MeV. The results should therefore be taken as
model-dependent measurements, which assume the Flatt\'e parametrization
with the parameter values of Zou and Bugg, and no interference
with background. Fits to the \pipi\ and \KK\ spectra are shown
in figures~\ref{xpbins} and~\ref{xpbinsphi} for six bins of $x_p$;
it can be seen from the figures that the parametrization 
works well in the Z$^0$ decay data over the entire kinematic
range.  

%\section{Results}
%\label{results}

\subsection{Fragmentation functions and total rates}
\label{fragf}

\begin{figure}[htbp]
\begin{center}
%\mbox{\epsfig{file=fragf.eps,height=17cm,width=17cm}}
\mbox{\epsfig{file=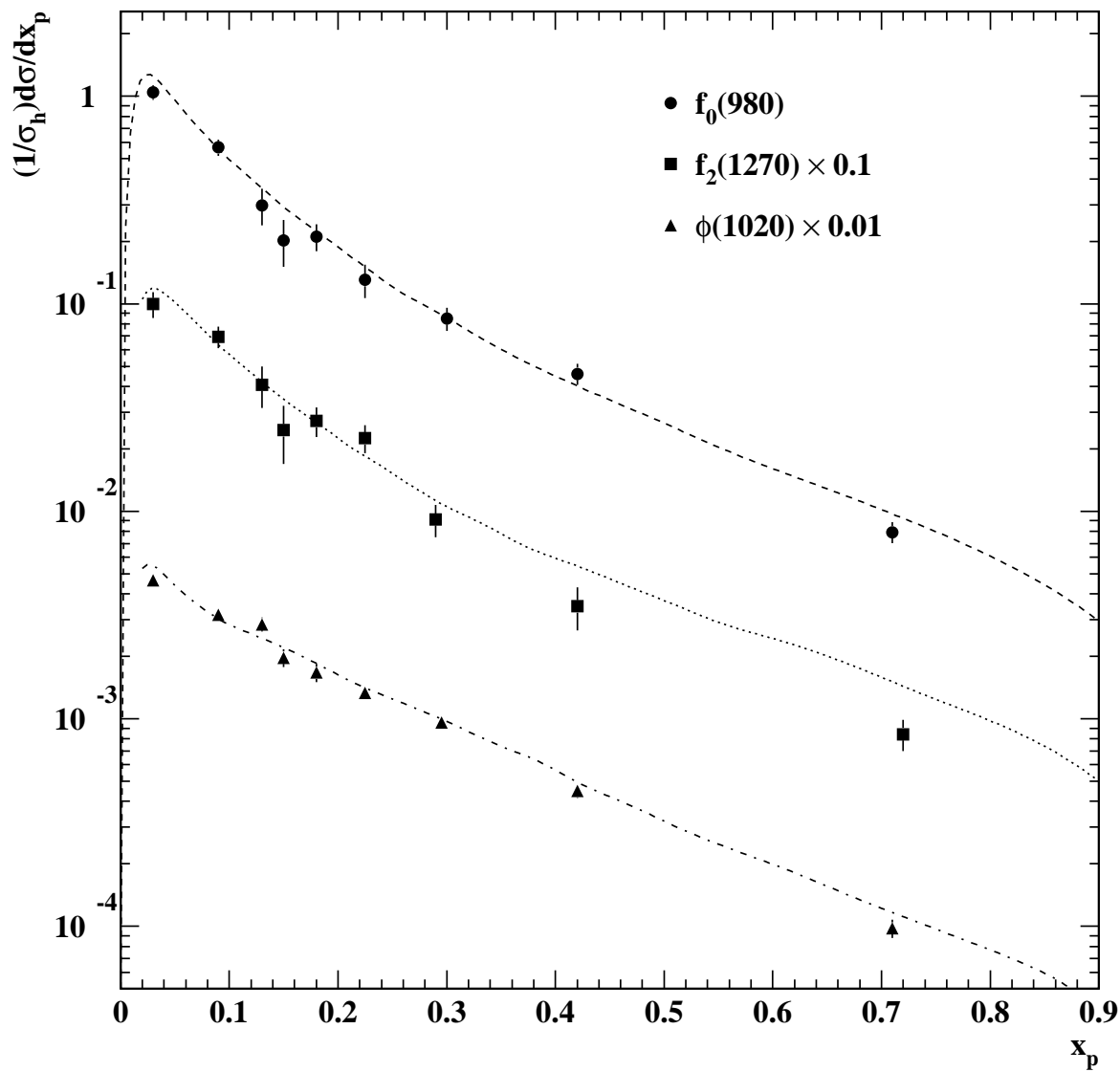,height=17cm,width=17cm}}
\end{center}
\caption{Measured fragmentation functions for \fz, \ft\ and $\phi$
together with the output of the JETSET 7.4 generator.  The \ft\ and
$\phi$ measurements have been scaled by $\times 0.1$ and $\times 0.01$
respectively. The Monte Carlo
curves have been normalized to the same total rate as in the data.
For \fz\ and \ft, the errors are statistical only. For the $\phi$, 
the uncorrelated systematic errors are included. 
There are additional, fully correlated, systematic errors
of $7.6\%$ for the \fz, $11.6\%$ for the \ft\ and
$3.5\%$ for the $\phi$.
The x-coordinates for plotting the data points were
evaluated following~\cite{stickem}.}
\label{fig-fragf}
\end{figure}

The rates obtained from the fits in bins of \xp\ are given in
table~\ref{tab-results}.
Figure~\ref{fig-fragf} shows these measured fragmentation functions,
$(1/\sigma_{\mathrm h})
{\mathrm d}\sigma/{\mathrm d}x_p$ (where $\sigma_{\mathrm h}$
is the total hadronic cross section),
along with curves from the 
JETSET 7.4 Monte Carlo generator, normalized for each
particle to the measured total rate seen in the data. 
%Although there are
%discrepancies between the data and the model for the overall rates, the
%shapes of the fragmentation functions are well modeled. Simple shifts 
%of the total intensities, 
%as could probably be obtained by re-tuning the model, would
%bring the curves and the measurements into good agreement. Since the 
%present tuning is based on rather limited knowledge of the production of
%${\mathrm L}=1$ mesons, discrepancies with data are not unexpected. 
Only statistical errors are shown for the \fz\ and \ft;  
the dominant systematic errors are 
correlated over the \xp\ bins and are discussed below in
section~\ref{systematics}. There is clearly good agreement for all
three particles between
the shapes of the momentum distributions in the Monte Carlo model and
those in the data. 

The total inclusive rates have been measured, by integrating the
fragmentation functions, to be: 
\begin{eqnarray}
0.141 \pm 0.007 \pm 0.011 & {\mathrm f}_0(980) \nonumber \\
0.155 \pm 0.011 \pm 0.018 & {\mathrm f}_2(1270) \nonumber \\ 
0.091 \pm 0.002 \pm 0.003 & \phi(1020)  \nonumber 
\end{eqnarray}
per hadronic Z$^0$ decay. The first errors are statistical
and the second systematic (discussed below). 
These rates compare with values
of $0.154 \pm 0.025$ \fz\ and
$0.240\pm0.061$ \ft\ mesons, obtained by B\"ohrer~\cite{Bohrer}
by extrapolating the DELPHI
measurements~\cite{Dlight} to the full range of \xp. The
measurement for the $\phi$ is in agreement with previous 
OPAL~\cite{OPALphi} and DELPHI~\cite{Dphi} results, but is
three standard deviations below an ALEPH~\cite{Aphi} measurement. 
The present $\phi$ measurement has the smallest statistical and 
systematic errors. 

\subsection{Systematic errors}
\label{systematics}

Table~\ref{tab-systematics} gives a summary of systematic errors on
the meson rates.
To estimate systematic errors for the \fz, the uncertainties in the
resonance mass and the couplings were considered.
The value of $m_0$ was varied from
0.951 to 0.963~GeV (the range obtained when the mass
was allowed to be free in the fits) while keeping the coupling
constants fixed. The maximum total variation, $2.5\%$,
was taken as the systematic error. Next $m_0$ and $g_{\mathrm K}$ were
fixed and the coupling $g_\pi$ was allowed to vary freely in the
fit. This gave a $7\%$ change in the
total rate, which was conservatively taken as the systematic error 
from this source. 
(If, instead, the errors in the $\pi\pi$ branching ratio from the
PDG are used, the systematic error would be $3\%$.) Finally, $m_0$
and $g_\pi$ were fixed and $g_{\mathrm K}$ was allowed to vary. The \fz\ rate 
changed by much less than the statistical error,
while the $\phi$ rate increased by $1.5\%$;
this was assigned as a systematic error on the $\phi$ measurement.
Systematic errors from uncertainties in Monte Carlo modelling of the 
track cuts contributes 1.4\%~\cite{OPALphi} for resonances decaying
to two charged particles. The overall systematic error
on the \fz\ measurements was therefore $7.6\%$.

For the \ft\ measurements, the main sources of systematic uncertainty
also come from the resonance parametrization, although in this 
case it is safe to assume a normal Breit-Wigner resonance. 
To take account of possible long tails in the 
line shape beyond the upper limit of the fits
to the mass spectra, the relativistic D-wave Breit-Wigner
used for the fits was integrated out to 2.2~GeV, five full widths above the
nominal peak position. This resulted in an increase of $30\%$ in the total
intensity over the fitted value.
Since the shape of such a resonance is uncertain
so far from the pole position, one half of this extra contribution 
was added to each measurement of the differential cross 
section (the results in table~\ref{tab-results} include these
corrections), and a systematic error of $30\%/\sqrt {12}$ was assigned. 
The fits were repeated, varying the \ft\ mass and width by one standard
deviation (using the PDG values) above and below their nominal values.
The maximum change in the measured 
rate, $6\%$, was assigned as a systematic error. 
To account for the high mass tail of the $\rho^0$ above 1.15~GeV, 
the fits were repeated assuming two extreme possibilities:
the default JETSET simulation, with the shape
truncated at 1.15~GeV; and a constant contribution set 
to the level in the simulation just below 1.15~GeV. The maximum
change, $3.2\%$, was taken as the systematic error.
With an error of $3.1\%$ on the branching ratio to \pipi\ and 
$1.4\%$ from the
modelling of the track cuts, the overall systematic
error on the \ft\ measurements was therefore $11.6\%$.

Systematic errors due to the simulation of the energy loss were
measured by varying the assumed mean values of the theoretical \dedx\ 
distributions for a given particle hypothesis, and the assumed 
resolution on the
energy loss measurements. Studies of well-identified
pions from K$^0_{\mathrm S}$ decays, protons from $\Lambda$ decays and
kaons from D$^0$ decays were used to place limits on the
maximum possible deviations of these quantities, and the analysis
was repeated, with the \dedx\ weights of the tracks being recalculated 
each time. Systematic errors were assigned as
the maximum measured deviations from the standard fit values.
The resulting errors were negligible for the \fz\ and \ft, and contributed
a $1.3\%$ error on the total $\phi$ rate. Since these errors varied with 
momentum, they are also given in table~\ref{tab-results} as
uncorrelated systematic errors.
Because the $\phi$ is close to threshold, there is a small background
due to conversion e$^+$e$^-$ pairs, particularly at low \xp\ (as seen 
in figure 3). In the
standard fit, this was included in the single background term,
$B_{\KK}$ of equation~(\ref{fitKK}).
When this component was allowed to vary independently, the total 
$\phi$ rate changed
by $2\%$, which was taken to be the systematic error from this effect. 
As in~\cite{OPALphi}, an error of $1\%$ comes from uncertainty in the
mass resolution and $1.4\%$ from modelling of the track cuts.
Finally an error of $1.2\%$ was included for the 
$\phi \rightarrow {\mathrm K^+}{\mathrm K^-}$ 
branching ratio, bringing the overall systematic
error to $3.5\%$ (including the contribution from variation of 
$g_{\mathrm K}$ discussed above). 

In all of the measurements, the possible presence of other f-meson
resonances (for example the f$_0(1360)$) in the $\pi^+\pi^-$ mass
spectrum has been neglected. While there is no evidence in the data
for such states, and therefore no need to include them
in the fits, their presence could in principle give rise to 
additional systematic errors, particularly for the f$_2$.

\subsection{Production as functions of event multiplicity and rapidity gap}
\label{sec-Gribov}

\begin{figure}[htbp]
\begin{center}
%\mbox{\epsfig{file=multbins.eps,height=17cm,width=17cm}}
\mbox{\epsfig{file=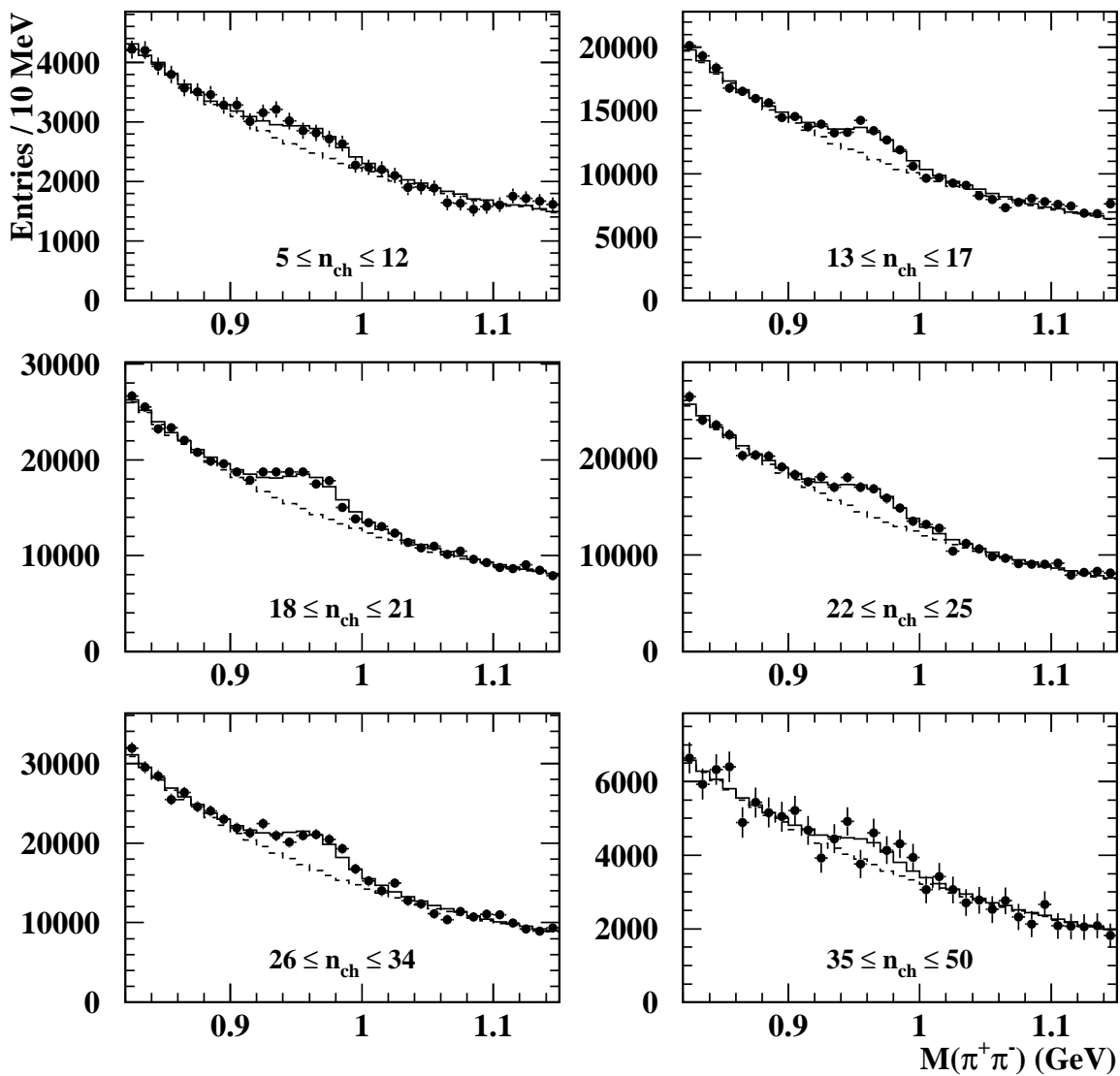,height=17cm,width=17cm}}
\end{center}
\caption{Fitted \pipi\ spectra (with like-sign spectra subtracted) in the \fz\
mass region for bins
of event charged-particle multiplicity. The points show 
the data, and the solid histograms give the results of the fits.
The dashed histograms show the fitted backgrounds.}
\label{multbins}
\end{figure}

\begin{figure}[htbp]
\begin{center}
%\mbox{\epsfig{file=multratios.eps,height=17cm,width=17cm}}
\mbox{\epsfig{file=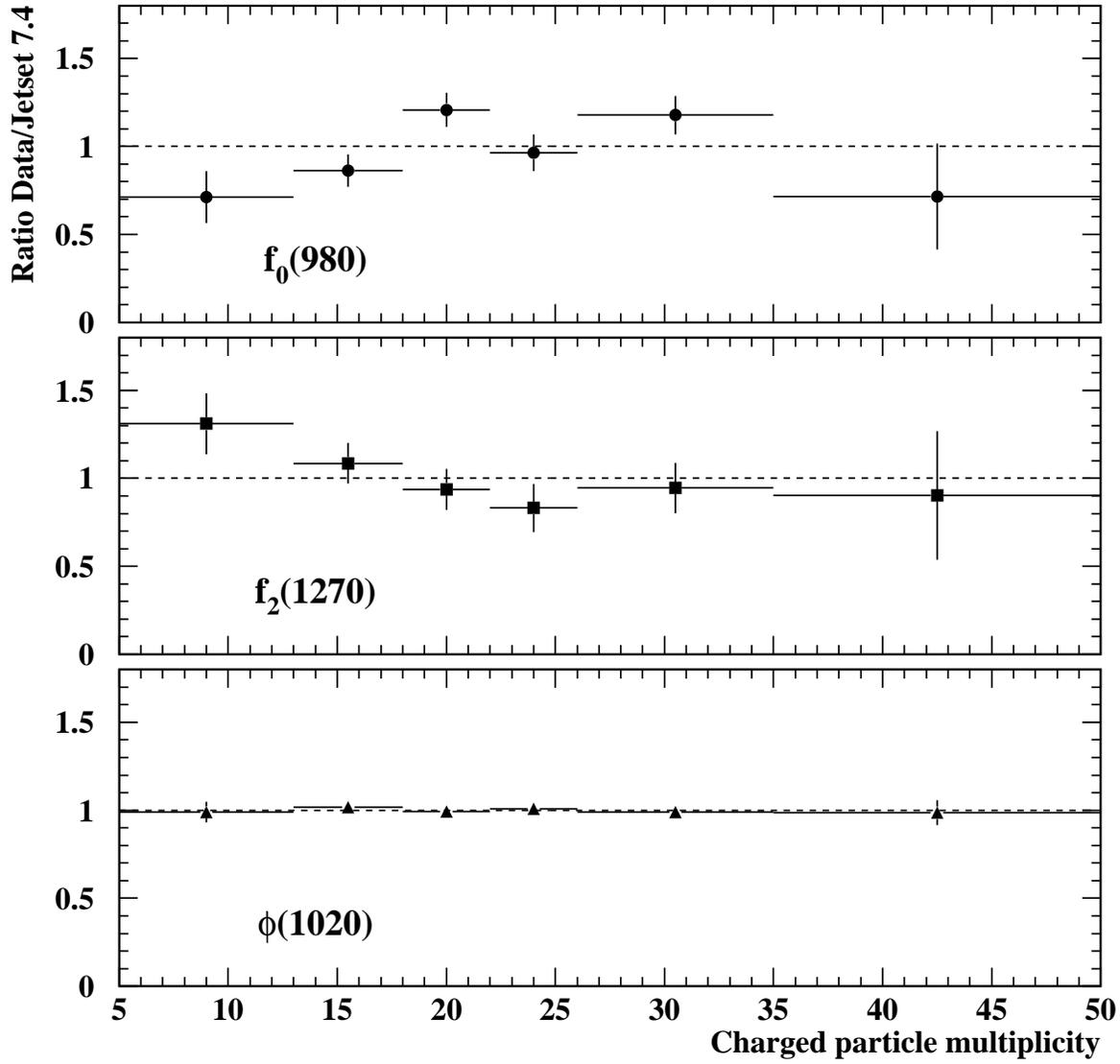,height=17cm,width=17cm}}
\end{center}
\caption{Ratios of production rates in data compared to the
detector-level JETSET 7.4 model, for bins of event charged-particle 
multiplicity. The errors are statistical only. 
The ratios have been normalized such that the weighted 
average is unity for each particle.}
\label{multratios}
\end{figure}

\begin{figure}[htbp]
\begin{center}
%\mbox{\epsfig{file=gapratios.eps,height=17cm,width=17cm}}
\mbox{\epsfig{file=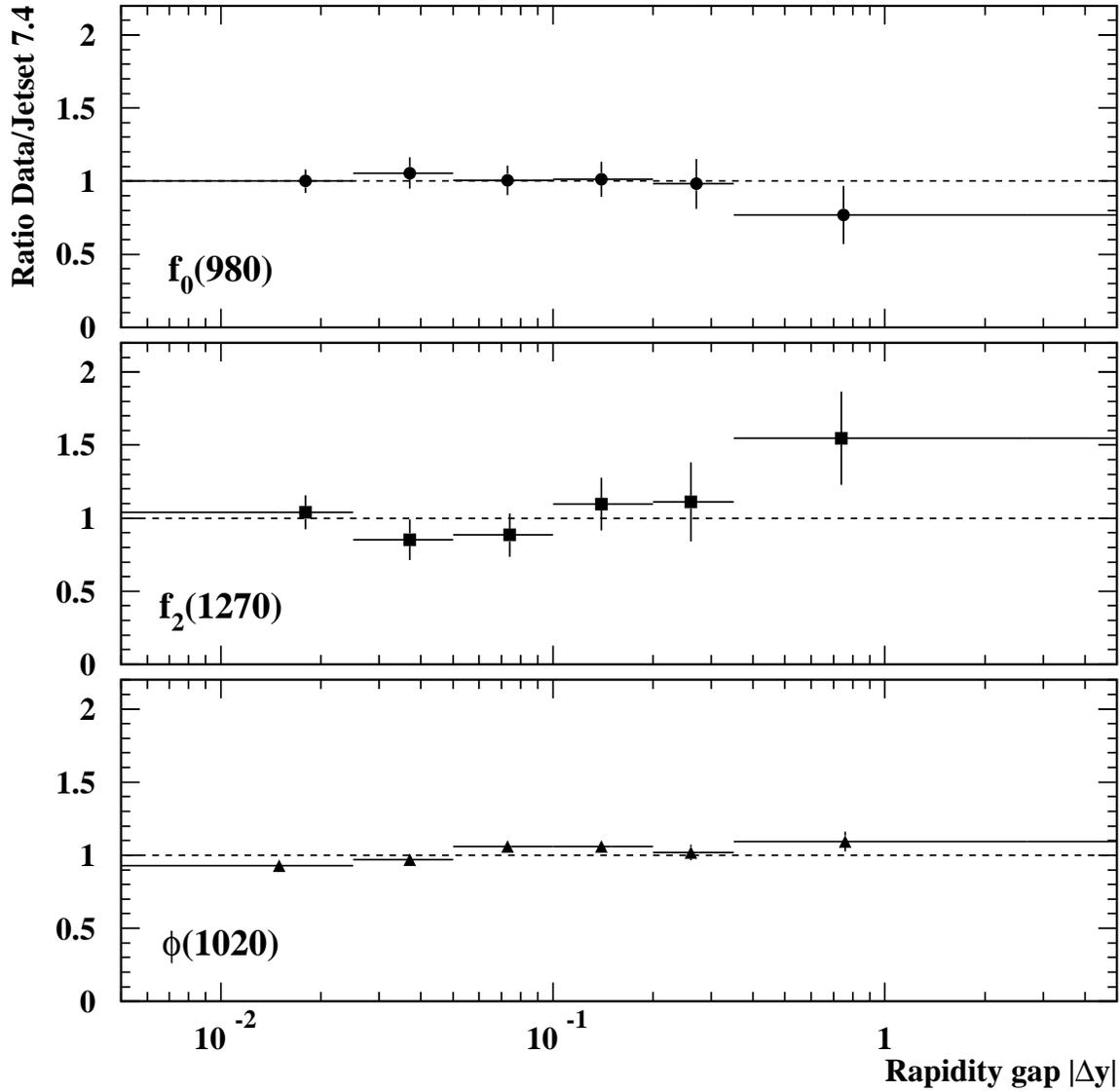,height=17cm,width=17cm}}
\end{center}
\caption{Ratios of total production rates in data compared to the
detector-level JETSET 7.4 model, for bins of the absolute value of 
the rapidity 
difference between the meson and the nearest charged particle.  
The errors are statistical only. 
The ratios have been normalized such that the weighted average 
is unity for each particle. The lowest bin extends down to zero, 
while the highest extends to the kinematic limit.}
\label{gapratios}
\end{figure}

As discussed in section~\ref{sec-intro}, the Gribov vacuum scalar
states are expected to be produced preferentially in low-multiplicity
events, and when isolated in rapidity relative to the other particles.
Figure~\ref{multbins} shows the fits to the
\pipi\ mass spectra in the \fz\ mass region for 
the six bins of event charged-track 
multiplicity. Figure~\ref{multratios} gives, 
for each of the three resonances, the ratios of production 
in data relative to the detector-level 
JETSET sample (in which the resonances are treated as
conventional mesons) as a function of the  
multiplicity. 
Figure~\ref{gapratios} shows the
same ratios as a function of the magnitude of the rapidity difference
between the resonances and the nearest charged particle.  
For each resonance the ratios have been normalized such that the weighted
averages are unity. 
No evidence is seen for anomalously large 
production of \fzz\ either at low multiplicity or at large rapidity gap,
with the Monte Carlo model giving a good description. 
The data therefore do not provide any evidence to support the 
hypothesis that the \fz\ is a manifestation
of a vacuum scalar state. 

\subsection{Production in quark and gluon jets}
\label{qg}

\begin{figure}[htbp]
\begin{center}
%\mbox{\epsfig{file=jetratios.eps,height=17cm,width=17cm}}
\mbox{\epsfig{file=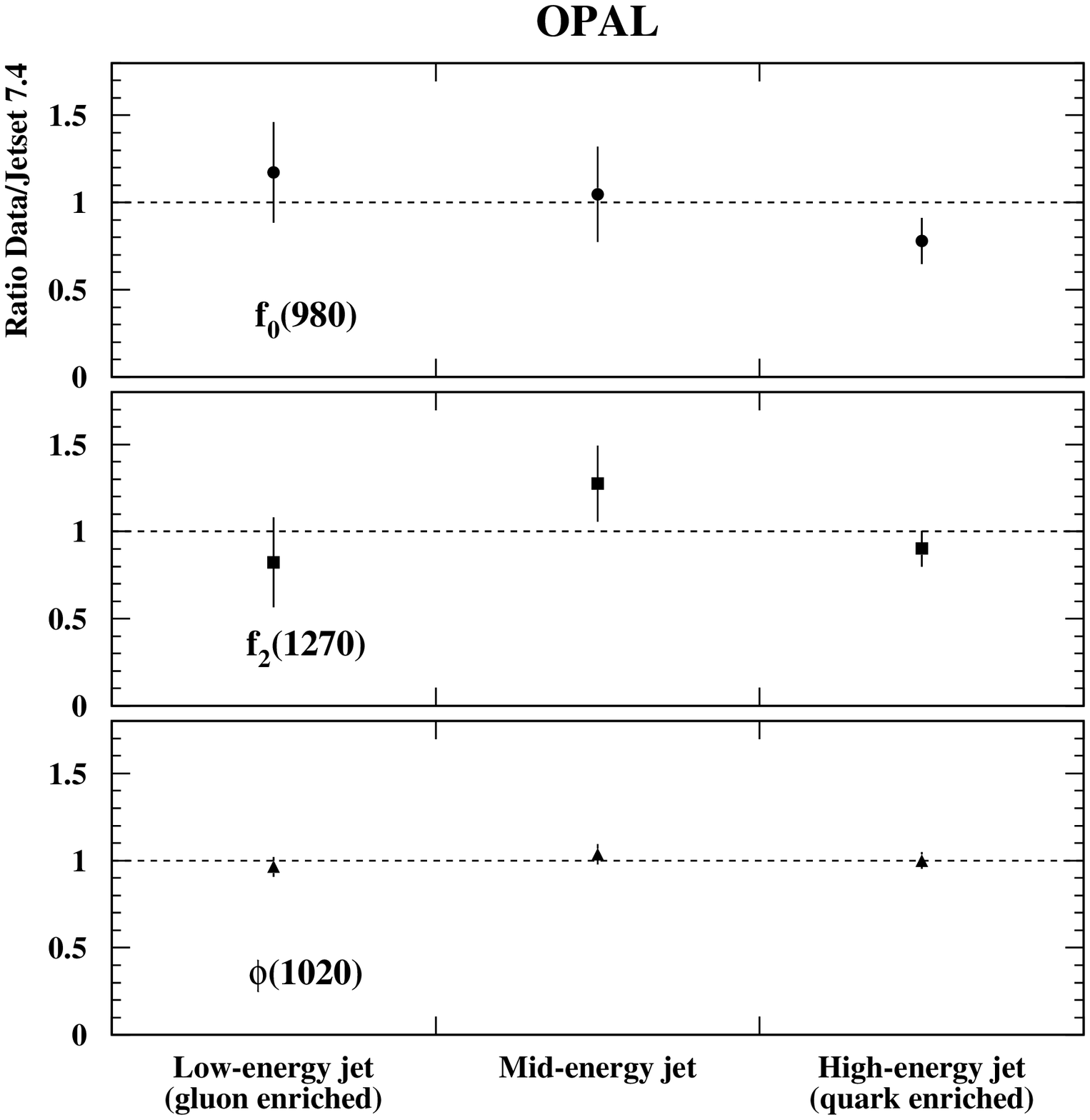,height=17cm,width=17cm}}
\end{center}
\caption{Ratios of total production rates in data compared to
the detector-level JETSET 7.4 model for the energy-ordered jets 
in three-jet events. The errors are statistical only. 
The ratios have been normalized for each particle 
to give a weighted average value of unity over
the three jets.}
\label{jetratios}
\end{figure}

The total rates in each of the jets in the selected three-jet events
were measured by fitting to the mass spectra in the three bins of 
scaled energy, and
summing the contributions. The jet-finding procedures described in
section~\ref{jetfinding} were also applied to the detector-level
Monte Carlo sample, and the total number of each of \fz, \ft\ and $\phi$
in each jet was counted. 
Figure~\ref{jetratios} shows the ratios of total rates measured in the
data relative to the detector-level Monte Carlo model for each of 
the three jet classes. The figure shows
no significant evidence for any differences between production in the
quark-enriched (high-energy) and the gluon-enriched (low-energy) jets.

\section{Summary and conclusions}
\label{sec-conclude}

Fragmentation functions and total inclusive rates 
in Z$^0$ decay have been measured
for three resonances, the \fzz, \ftwo and $\phi(1020)$, using the
full LEP 1 statistics of OPAL. The \fz\ and \ft\ measurements will provide
input to understanding the physics of inclusive particle production
in the P-wave meson sector, particularly in the context of the 
Monte Carlo models. 

The production characteristics of the \fz\ show no significant 
differences
from those of the \ft\ and $\phi$ mesons. In particular, the shapes
of the fragmentation functions are similar for all three, and are
well reproduced by the JETSET 7.4 Monte Carlo model, within which 
the \fz\ is
treated as a conventional scalar meson. 
The total inclusive rate for the \fz\ is $0.141 \pm 0.013$ per
hadronic Z$^0$ decay,
similar to the value, $0.155 \pm 0.021$, measured for the \ft.
The \fz\ rate is significantly larger 
than for the $\phi$, $0.091 \pm 0.004$ per hadronic Z$^0$ decay. 
These features are 
consistent with the \fz\ being, like the \ft, a \qqbar\ meson 
in the $^3$P state composed mainly of u$\overline {\mathrm u}$ and 
d$\overline {\mathrm d}$.
  
For all three states, the production rates relative to the Monte
Carlo model have been measured 
as functions of the charged particle multiplicity of
the event and the gap in rapidity to the nearest charged particle.
The distributions are found to be flat in all cases.
In particular, no evidence has been found for enhanced \fz\ 
production at low multiplicities or at large rapidity gap. 
There is thus no evidence to identify the \fz\ with the vacuum scalar
state proposed by Gribov.

Production in energy-ordered jets in three-jet events has been
measured with a view to seeking differences between quark and
gluon induced jets. No significant differences are seen
between the data and the Monte Carlo model in the relative 
production rates. There is therefore
no evidence for any enhanced gluon content in the \fz.  

In summary, all measured characteristics of \fzz\ production in the
Z$^0$ decay data of OPAL are consistent with its interpretation as a
conventional scalar meson. Quantitative theoretical or phenomenological 
predictions for
production of the types of states discussed in the 
introduction could enable more definite conclusions to be drawn from
the Z$^0$ data.  

\bigskip
\bigskip
\noindent
{\bf Acknowledgements}

%\noindent
%The usual acknowledgements ...
%We would like to thank Sandy Donnachie and Orlando Villalobos-Baillie 
%for discussions about the \fzz. 
%\bigskip\bigskip\bigskip
%\appendix
\par
%Acknowledgements:
%\par
\noindent
We particularly wish to thank the SL Division for the efficient operation
of the LEP accelerator at all energies
 and for
their continuing close cooperation with
our experimental group.  We thank our colleagues from CEA, DAPNIA/SPP,
CE-Saclay for their efforts over the years on the time-of-flight and trigger
systems which we continue to use.  In addition to the support staff at our own
institutions we are pleased to acknowledge the  \\
Department of Energy, USA, \\
National Science Foundation, USA, \\
Particle Physics and Astronomy Research Council, UK, \\
Natural Sciences and Engineering Research Council, Canada, \\
Israel Science Foundation, administered by the Israel
Academy of Science and Humanities, \\
Minerva Gesellschaft, \\
Benoziyo Center for High Energy Physics,\\
Japanese Ministry of Education, Science and Culture (the
Monbusho) and a grant under the Monbusho International
Science Research Program,\\
German Israeli Bi-national Science Foundation (GIF), \\
Bundesministerium f\"ur Bildung, Wissenschaft,
Forschung und Technologie, Germany, \\
National Research Council of Canada, \\
Research Corporation, USA,\\
Hungarian Foundation for Scientific Research, OTKA T-016660, 
T023793 and OTKA F-023259.\\

\vfill

\newpage
\begin{table}[htb]
\begin{center}
\begin{tabular}{|r@{--}l|r@{$\pm$}l|r@{$\pm$}l|r@{$\pm$}c@{$\pm$}l|}
\hline
\multicolumn{2}{|c|}{ \xp\ range} & \multicolumn{2}{c|}{ \fzz\ } & 
\multicolumn{2}{c|}{ \ftwo\ } & \multicolumn{3}{c|}{ $\phi(1020)$ } \\
\hline
0.00&0.06 &1.04&0.09     &1.00&0.14   &0.464&0.011&0.005   \\
0.06&0.12 &0.57&0.05     &0.69&0.08   &0.316&0.021&0.007 \\
0.12&0.14 &0.30&0.06     &0.41&0.09   &0.285&0.020&0.009   \\
0.14&0.16 &0.20&0.05     &0.25&0.08   &0.197&0.019&0.006  \\
0.16&0.20 &0.21&0.03     &0.27&0.04   &0.167&0.017&0.002   \\
0.20&0.25 &0.13&0.02     &0.22&0.03   &0.133&0.007&0.002   \\
0.25&0.35 &0.085&0.011   &0.091&0.016 &0.096&0.004&0.001 \\
0.35&0.50 &0.046&0.005   &0.035&0.008 &0.045&0.002&0.001 \\
0.50&1.00 &0.0079&0.0009 &0.008&0.001 &0.010&0.001&0.000 \\
\hline
0.00&1.00 &0.141&0.007   &0.155&0.011 &0.091&0.002&0.001 \\
\hline
\end{tabular}
\end{center}
\caption{Measured differential cross sections, 
$(1/\sigma_{\mathrm h}){\mathrm d}\sigma/{\mathrm d}x_p$, and 
total inclusive rates
for \fz, \ft\ and $\phi$
production. There are additional overall systematic errors
of $7.6\%$ for the \fz, $11.6\%$ for the \ft\ and
$3.5\%$ for the $\phi$. The uncorrelated systematic errors for 
the $\phi$ are given in the table; the error for the $x_p$
range $0.50-1.00$ is smaller than $0.0005$. 
The uncorrelated systematic errors for the \fz\ and \ft\
are negligible in comparison with the statistical errors and the 
correlated systematic errors.}
\label{tab-results}
\end{table}

\begin{table}[htb]
\begin{center}
\begin{tabular}{|l|c|c|c|}
\hline
       &  \fzz\  & \ftwo\  & $\phi(1020)$ \\
\hline
Variation of $m_0$           & 2.5\% &  --    &  --    \\
Variation of $g_\pi$         & 7.0\% &  --    &  --    \\ 
Variation of $g_{\mathrm K}$ &  --    &  --    & 1.5\% \\
f$_2$ resonance line shape   &  --    & 8.7\% &  --    \\
f$_2$ mass and width         &  --    & 6.0\% &  --    \\
BR(f$_2 \rightarrow \pi^+\pi^-$)     
                             &  --    & 3.1\% &  --    \\
$\rho^0$ resonance line shape&  --    & 3.2\% &  --    \\
d$E$/d$x$ parametrization    &  -    &  -    & 1.3\% \\
BR($\phi \rightarrow {\mathrm K}^+{\mathrm K}^-$)    
                             &  --    &  --    & 1.2\% \\
K$^+$K$^-$ mass resolution   &  --    &  --    & 1.0\% \\
Photon conversions           & -- & -- & 2.0\% \\
Track cuts                   & 1.4\% & 1.4\% & 1.4\% \\
\hline
Total                        & 7.6\% & 11.6\%& 3.5\% \\
\hline
\end{tabular}
\end{center}
\caption{Sources of systematic errors on the total inclusive 
meson rates. Errors on the $\phi$ rates from uncertainties in the 
parametrization of d$E$/d$x$ are momentum dependent, and 
are given in table~1 for the bins of $x_p$.}
\label{tab-systematics}
\end{table}

\newpage

\newpage

\newpage

\newpage

\newpage

\newpage

\newpage

\newpage


\newpage
\begin{thebibliography}{99}

\bibitem{Bohrer} A. B\"ohrer, Phys. Rep. 291 (1997) 107. 
\bibitem{Pennington} M.~R.~Pennigton, in Proceedings of the 6th
International Conference on Hadron Spectroscopy, Ed. M.~C.~Birse,
G.~D.~Lafferty and J.~A.~McGovern, World Scientific (1996).
\bibitem{PDG} Particle Data Group, R. M. Barnett al.,
Phys. Rev. D54 (1996) 1.
\bibitem{Jaffe} R.~L.~Jaffe and K.~Johnson, Phys. Lett. B60 (1976) 201; \\
R.~L.~Jaffe, Phys. Rev. D15 (1977) 267 and Phys. Rev. D15 (1977) 269.
\bibitem{Weinstein} J.~Weinstein and N.~Isgur, Phys. Rev. D41 (1990) 2236.
\bibitem{Gribov} V. N. Gribov, Lund preprint, LU-TP 91-7.
\bibitem{Close} F. E. Close et al., Phys. Lett. B319 (1993) 291.
\bibitem{Ishida} S. Ishida et al., in Proceedings of the 6th
International Conference on Hadron Spectroscopy, Ed. M.~C.~Birse,
G.~D.~Lafferty and J.~A.~McGovern, World Scientific (1996); \\
S.~Ishida, H.~Sawazaki, M.~Oda and K.~Yamada, Phys. Rev. D47 (1993) 179.
\bibitem{Robson} D.~Robson, N. Phys. B130 (1977) 328.
\bibitem{Closeglue} C.~Amsler and F.~E.~Close, Phys. Lett. B353 (1995) 385.
\bibitem{Lattice} G. Bali et al., Phys. Lett. B307 (1993) 378; \\
D. Weingarten et al., N. Phys. B34 (1994) 29.
\bibitem{Tornqvist} N.~A.~T\"ornqvist et al., Z. Phys. C68 (1995) 647.
\bibitem{Bugg} B.~S.~Zou and D.~V.~Bugg, Phys. Rev. D48 (1993) R3948.
\bibitem{Anisovich} A.~V.~Anisovich and A.~V.~Sarantsev, 
Phys. Lett. B413 (1997) 137.
\bibitem{JETSET} T. Sj\"{o}strand, Comp. Phys. Commun. {82} (1994) 74.
\bibitem{string} B. Andersson, G. Gustafson, G. Ingelman 
and T. Sj\"ostrand, Phys. Rep. {97} (1983) 31.
\bibitem{toprev} I.~G.~Knowles and G.~D.~Lafferty, 
J.~Phys.~G: Nucl.~Part.~Phys. {23} (1997) 731.
%\bibitem{BECs} ALEPH Coll., D.~Buskulic et al., Z. Phys. C64 (1994) 361; \\
%DELPHI Coll., P. Abreu et al., Phys. Lett. B379 (1996) 330; \\
%OPAL Coll., R. Akers et al., Z. Phys. C67 (1995) 389.
\bibitem{Dlight} DELPHI Coll., P. Abreu et al., Z. Phys. C65 (1994) 587.
\bibitem{Arho} ALEPH Coll., D. Buskulic et al., Z. Phys. C69 (1995) 379.
\bibitem{OPALdet} OPAL Coll., K. Ahmet et al., Nucl. Instrum. 
Methods  {A305} (1991) 275.
\bibitem{silicon} P.P.~Allport et al., 
Nucl. Instrum. Methods  {A324} (1993) 34; \\
P.P.~Allport et al., Nucl. Instrum. Methods  {A346} (1994) 476.
\bibitem{DEDX} M. Hauschild et al., Nucl. Instrum. Methods  {A314} (1992) 74.
\bibitem{GOPAL} J. Allison et al., Nucl. Instrum. Methods  {A317} (1992) 47.
\bibitem{JTtune} OPAL Coll., M.Z.~Akrawy et al., Z.~Phys. {C47} (1990) 505; \\
  The JETSET 7.4 parameters were tuned as described in 
  OPAL Coll., G.~Alexander et al., Z. Phys.  {C69} (1996) 543.
\bibitem{TKMH} OPAL Coll., G. Alexander at al., Z. Phys.  {C5} (1991) 175.
\bibitem{flatte} J.~B.~Gay~et al., Phys. Lett. 63B (1976) 220; \\
S. M. Flatt\'e, Phys. Lett. 63B (1976) 224 and 
Phys. Lett. 63B (1976) 228.
\bibitem{Durham} S. Catani, Yu.~L.~Dokshitzer, F.~Fiorani and
B.~R.~Webber, N.~Phys.~B377~(1992)~445; \\
S. Catani et al., Phys. Lett. B269 (1991) 432; \\
S. Bethke, Z. Kunszt, D.~E.~Soper and W.~J.~Stirling, 
N.~Phys.~B370~(1992)~310; \\
N. Brown and W. J. Stirling, Z. Phys. C53 (1992) 629.
% \bibitem{massless} OPAL Coll., P.~D.~Acton et al., Z.~Phys.~C54 (1992) 193.
% \bibitem{Bonn} ``Identified charged particle production in quark and
% gluon jets'', OPAL Physics Note PN299, submitted to the International
% Europhysics Conference on High Energy Physics, 
% Jerusalem, 19-26 August 1997.
% OPAL Coll., ``$\Lambda$ and K$^0_{\mathrm S}$ production in
% quark and gluon jets classified by energy ordering'', OPAL Physics Note
% PN236 (1996), submitted to XXVIII International Conference on High
% Energy Physics, July 1996, Warsaw.
\bibitem{OPALphi} OPAL Coll., R. Akers et al., Z.~Phys.~C68 (1995) 1.
\bibitem{Aphi} ALEPH Coll., D. Buskulic et al., Z. Phys. C69 (1996) 379.
\bibitem{Dphi} DELPHI Coll., P. Abreu et al., Z. Phys. C73 (1996) 61.
\bibitem{stickem} G. D. Lafferty and T. R. Wyatt, Nucl. Instrum. Methods
A355~(1995)~541.
\end{thebibliography}
\end{document}